\documentclass[aps,prb,twocolumn,10pt,superscriptaddress,showpacs]{revtex4-1}

\usepackage{graphicx}
\usepackage{dcolumn}
\usepackage{bm}

\begin{document}


\title{Evidence for a Ru$^{4+}$ Kondo Lattice in LaCu$_3$Ru$_4$O$_{12}$}


\author{S. Riegg
\affiliation{Experimental Physics V, Center for Electronic Correlations and Magnetism, University of Augsburg, D-86159 Augsburg, Germany}
}
\author{S. Widmann}
\affiliation{Experimental Physics V, Center for Electronic Correlations and Magnetism, University of Augsburg, D-86159 Augsburg, Germany}

\author{B. Meir}
\affiliation{Experimental Physics V, Center for Electronic Correlations and Magnetism, University of Augsburg, D-86159 Augsburg, Germany}

\author{S. Sterz}
\affiliation{Experimental Physics V, Center for Electronic Correlations and Magnetism, University of Augsburg, D-86159 Augsburg, Germany}

\author{A. G\"unther}
\affiliation{Experimental Physics V, Center for Electronic Correlations and Magnetism, University of Augsburg, D-86159 Augsburg, Germany}

\author{N. B\"uttgen}
\affiliation{Experimental Physics V, Center for Electronic Correlations and Magnetism, University of Augsburg, D-86159 Augsburg, Germany}

\author{S. G. Ebbinghaus}
\affiliation{Solid State Chemistry, Martin-Luther University Halle-Wittenberg, D-06099 Halle, Germany}

\author{A. Reller}
\affiliation{Resource Strategy, University of Augsburg, D-86159 Augsburg, Germany}

\author{H.-A. Krug von Nidda}
\email[]{hans-albrecht.krug@physik.uni-augsburg.de}
\affiliation{Experimental Physics V, Center for Electronic Correlations and Magnetism, University of Augsburg, D-86159 Augsburg, Germany}

\author{A. Loidl}
\affiliation{Experimental Physics V, Center for Electronic Correlations and Magnetism, University of Augsburg, D-86159 Augsburg, Germany}



\date{\today}

\begin{abstract}
Rare $d$-electron derived heavy-fermion properties of the solid-solution series LaCu$_3$Ru$_x$Ti$_{4-x}$O$_{12}$ were studied for $1 \leq x \leq 4$ by resistivity, susceptibility, specific-heat measurements, and magnetic-resonance techniques. The pure ruthenate ($x = 4$) is a heavy-fermion metal characterized by a resistivity proportional to $T^2$ at low temperatures $T$. The coherent Kondo lattice formed by the localized Ru 4$d$ electrons is screened by the conduction electrons leading to strongly enhanced effective electron masses. By increasing titanium substitution the Kondo lattice becomes diluted resulting in single-ion Kondo properties like in the paradigm $4f$-based heavy-fermion compound Ce$_x$La$_{1-x}$Cu$_{2.05}$Si$_2$ [M. Ocko {\em et al.}, Phys. Rev. B \textbf{64}, 195106 (2001)]. In LaCu$_3$Ru$_x$Ti$_{4-x}$O$_{12}$ the heavy-fermion behavior finally breaks down on crossing the metal-to-insulator transition close to $x = 2$.

\end{abstract}

%

\maketitle


\section{Introduction}
The discovery of heavy-fermion properties in the transition-metal oxide LiV$_2$O$_4$ in 1997 opened new horizons\cite{Kond97,Kri99} in the field of strongly correlated electron systems, because to that date the term "heavy fermions" was confined to intermetallic compounds containing elements with partially filled $4f$ or $5f$ shells, like Ce, Yb or U, as there are e.g. CeAl$_3$, CeCu$_2$Si$_2$, YbRh$_2$Si$_2$, and UPt$_3$ (Refs. \onlinecite{And75,Ste79,Tro00,Ste84}). In those materials, below a certain characteristic temperature $T^*$ of the order of 10\,K, the local $f$ moments become screened by the conduction electrons due to the Kondo effect.\cite{Kond64} This gives rise to metallic properties with strongly enhanced effective masses $m^* \approx 10^3 m_{\rm e}$ with respect to the free electron mass $m_{\rm e}$ resulting, e.g., in an enhanced Sommerfeld coefficient $\gamma \propto m^*$ of the specific heat as well as in an enhanced Pauli susceptibility $\chi \propto m^*$ as a consequence of the enhanced density of states at the Fermi energy $N(E_{\rm F}) \propto m^*$. Later on, in LiV$_2$O$_4$ Kondo screening as origin of heavy-fermion behavior has been questioned,\cite{Fulde01} and enhanced effective masses have been explained by frustration effects. And indeed, the $B$ site of spinels $AB_2$O$_4$ forms a pyrochlore lattice, one of the strongest contenders of magnetic frustration. In the aftermath enlarged effective masses have been also detected in several titanates \cite{Tok93,Tag93,Hei01} and ruthenates.\cite{Cao97,Mae94,Mae01,Gri01,Kha01,Tan09} Again, in these transition-metal oxides other mechanisms than the Kondo-effect, such as frustration or the proximity to a metal-to-insulator transition (MIT), are inferred as reasons for spin fluctuations and concomitant electronic mass enhancement.\cite{Krug2003} Note in the latter case, spin fluctuations and effective masses strongly increase on approaching the MIT from the metallic side.\cite{Krug2003}

In the group of ruthenium based heavy fermions, the perovskite derived $A$Cu$_3$Ru$_4$O$_{12}$ ($A=$ Na, Ca, Sr, La, Pr, Nd) deserve special attention. \cite{Ebb02,Hebert15} Detailed investigations revealed CaCu$_3$Ru$_4$O$_{12}$ as moderately mass enhanced Fermi-liquid with a Sommerfeld coefficient $\gamma = 85$~mJ/(mol K$^{2}$) showing intermediate valence properties above 2~K, but non-Fermi-liquid behavior for lower temperatures.\cite{Kob04,Kri08,Kri09} In search for a quantum-critical point, substitution of Ru by non-magnetic Ti$^{4+}$ (3$d^0$) was expected to drive the system through a MIT up to the antiferromagnetic Mott insulator CaCu$_3$Ti$_4$O$_{12}$ famous for its colossal dielectric constant.\cite{Hom01,Lun04} However, it is impossible to synthesize single-phase samples of CaCu$_3$Ru$_x$Ti$_{4-x}$O$_{12}$ in the range $1.5 < x < 4$ due to a miscibility gap.\cite{Ram04,Kob04,Tsu09} In contrast, the isostructural lanthanum compound LaCu$_3$Ru$_4$O$_{12}$ with a comparably large Sommerfeld coefficient of 139~mJ/(mol K$^2$) does not exhibit any miscibility gap on substitution by titanium and, thus, LaCu$_3$Ru$_x$Ti$_{4-x}$O$_{12}$ turns out to be well suited to study the MIT. Previous investigations of magnetic and transport properties revealed semiconducting behavior with a spin-glass ground state on the Ti rich side $(x<2)$, but a purely metallic paramagnet on the Ru rich side $(x>3)$ (Ref. \onlinecite{Buet10}). The exact concentration $x_c$ of the MIT could not be detected accurately by resistivity measurements,\cite{Ebb10} but was later on localized by means of electron spin resonance slightly above $x=2$ (Ref. \onlinecite{Schmi14}). The full phase diagram from antiferromagnetic insulator ($x=0$) to heavy-fermion metal ($x=2$) has been documented recently in Ref.~\onlinecite{Riegg15a}. In the present paper, we concentrate on the metallic regime $2 < x < 4$ to investigate the Fermi-liquid properties on approaching the MIT by means of combined susceptibility, resistivity, specific heat, electron spin resonance (ESR), and nuclear magnetic resonance (NMR and NQR) measurements. We will provide evidence that in LaCu$_3$Ru$_4$O$_{12}$, in contrast to previously reported $d$-electron derived heavy-fermion compounds, the Kondo-like scattering of conduction electrons at the Ru site can be identified as the most reasonable origin of Ru spin fluctuations. This Kondo-like behavior especially manifests in the local electronic properties probed by the magnetic resonance techniques.

\section{Experimental Details}
Polycrystalline samples of LaCu$_3$Ru$_x$Ti$_{4-x}$O$_{12}$ were synthesized by solid-state reaction utilizing the binary metal oxides.\cite{Riegg15}
X-ray diffraction confirmed the proper cubic phase with the lattice parameter $a$ monotonously increasing with increasing $x$. Magnetic susceptibilities were measured in the temperature range $1.8 \leq T \leq 400$~K using a superconducting-quantum-interference-device (SQUID) magnetometer MPMS5-XL (\textit{Quantum Design}). Specific-heat and electrical-resistivity measurements were performed by means of a physical-property-measurement-system PPMS (\textit{Quantum Design}) allowing for temperatures $1.8 \leq T \leq 300$~K (Ref. \onlinecite{rho}).

For nuclear magnetic resonance (NMR and NQR) we employed a home built setup.\cite{NQR} For the isotopes $^{99}$Ru, $^{139}$La,  and $^{63}$Cu we obtained the local susceptibility via temperature dependent NMR line-shift measurements.  In addition, the $^{63}$Cu nuclei served as intrinsic nuclear-spin probe $(I=3/2)$ in pure quadrupole resonance NQR measurements of the spin-lattice relaxation rate $1/T_1$. These NQR measurements take place in zero applied magnetic fields and exploit the interaction between the nuclear quadrupole moment and the surrounding electrical field gradient for lifting the nuclear spin degeneracy. Electron spin resonance (ESR) spectra were measured at constant microwave frequency of $\nu \approx 9.4$~GHz (X-band) as function of a static external magnetic field using a \textit{Bruker} ELEXSYS E500 CW spectrometer equipped with a continuous He-gas flow cryostat (\textit{Oxford Instruments}) working in the temperature range from liquid helium to room temperature.\cite{ESR} As for $x > 0$ the spins of both ruthenium and copper relax too fast to yield a detectable ESR signal, the samples were doped with 5\% of gadolinium on the lanthanum site. Trivalent Gd$^{3+}$ (electronic configuration $4f^7$, spin $S = 7/2$, and angular momentum $L = 0$) has been shown to be an appropriate ESR probe in metals to detect the local electronic properties.\cite{Buet97,KvN98}

\section{Sample Characterization}

\begin{figure}
\includegraphics[width=0.4\textwidth]{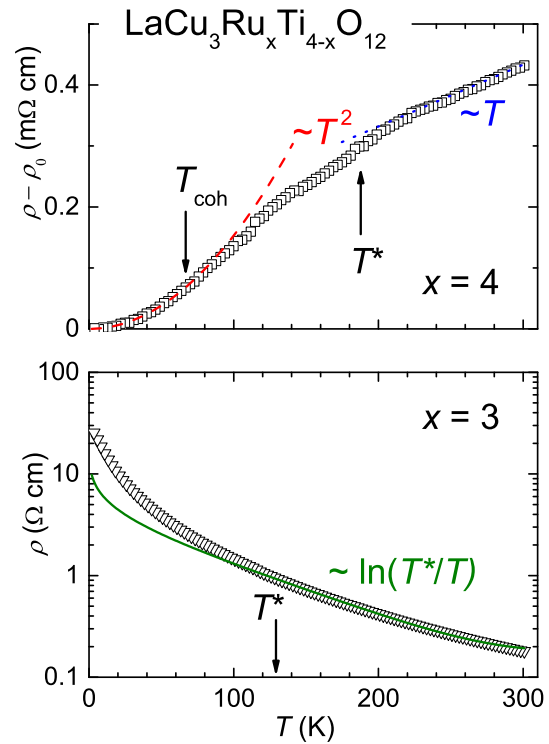}%
\caption{(Color online) Resistivity of LaCu$_3$Ru$_x$Ti$_{4-x}$O$_{12}$ for $x = 4$ (upper frame) and 3 (lower frame). For $x=4$ a residual resistivity of $\rho_0 = 3 \times 10^{-4} \Omega$cm has been subtracted. $T_{\rm coh}$ and $T^*$ denote formation of the coherent Kondo-lattice ground state and the onset of moment compensation, respectively. The power laws $\propto T^2$ (dashed) and $\propto T$ (dotted) indicate dominance of electron-electron and electron-phonon scattering, respectively. For $x=3$ the solid line is a fit by $\rho=\rho_0+a \cdot T + d \cdot \ln(T^*/T)$. \label{Ru_x3and4}}
\end{figure}

Fig.~\ref{Ru_x3and4} exemplarily illustrates the temperature dependence of the electrical resistivity $\rho(T)$ for two selected samples of the metallic regime ($x > 2$), i.e. LaCu$_3$Ru$_4$O$_{12}$ ($x=4$, upper frame) and LaCu$_3$Ru$_3$TiO$_{12}$ ($x=3$, lower frame). For $x=4$, as expected for a metallic sample, the resistivity increases monotonously with temperature. At elevated temperatures, $T>200$~K, the linear dependence typical for dominant electron-phonon scattering shows up. In contrast at low temperatures, $T<50$\,K, the data are well described by a quadratic dependence $\rho = \rho_0 + A T^2$, which is typical for heavy-fermion metals, where the contribution of electron-electron scattering $A \propto (m^*)^2$ becomes sizable because of strongly enhanced effective electronic masses $m^*$. In simple metals this behavior is usually masked by the low-temperature phonon-scattering contribution $\rho_{\rm ph} \propto T^5$. For LaCu$_3$Ru$_4$O$_{12}$ a value of $A=15.4$~n$\rm \Omega$cm/K$^2$ is obtained, \cite{Buet10} which is comparable with the result published by Tanaka \textit{et al.} (Ref. \onlinecite{Tan09}). Such a behavior is very similar to that of $4f$-based Kondo-lattice compounds like CeSn$_3$ (Ref. \onlinecite{Sereni80, Umehara90}) or Ce(Pt$_{1-x}$Ni$_x$)$_2$Si$_2$ ($x \geq 0.6$, Ref. \onlinecite{Ragel04}) close to intermediate valence. Those heavy-fermion compounds with moderately enhanced effective electron masses are characterized by a characteristic temperature $T^* \sim 200$\,K (Kondo temperature), where the resistivity deviates from the linear temperature dependence and a so called coherence temperature $T_{\rm coh} \sim 50$\,K, which denotes the upper boundary of the coherent Fermi-liquid regime with $\rho \propto T^2$.
This strong similarity suggests to use the identical notation ($T^*, T_{\rm coh}$) for the respective temperature regimes in LaCu$_3$Ru$_4$O$_{12}$.

For $x=3$, at first glance the monotonous increase of the resistivity on decreasing temperature seems to be in contradiction with metallic properties. But the temperature dependence does not resemble semiconducting behavior either. The resistivity increases only by two orders of magnitude when lowering the temperature from 300~K to 2~K. On further inspection the data follow a logarithmic dependence $\rho \propto ln(T^*/T)$ with $T^* \approx 130$~K in a wide temperature range $T > 80$~K. This is comparable to the scenario observed on dilution of dense $4f$-electron based Kondo-lattice compounds evolving from a coherent Fermi-liquid state into the single-ion Kondo effect like, e.g., reported for Ce$_x$La$_{1-x}$Cu$_{2.05}$Si$_2$ (Ref. \onlinecite{Ock01}). This means that $T^*$ can be identified with the single-ion Kondo temperature. To lower temperatures $T < 80$~K the experimental resistivity of LaCu$_3$Ru$_3$TiO$_{12}$ increases faster than the fit curve, but still qualitatively exhibits a quite similar behavior. This may indicate a distribution of Kondo temperatures due to the statistic substitution of the ruthenium ions by titanium.

\begin{figure}
\includegraphics[width=0.4\textwidth]{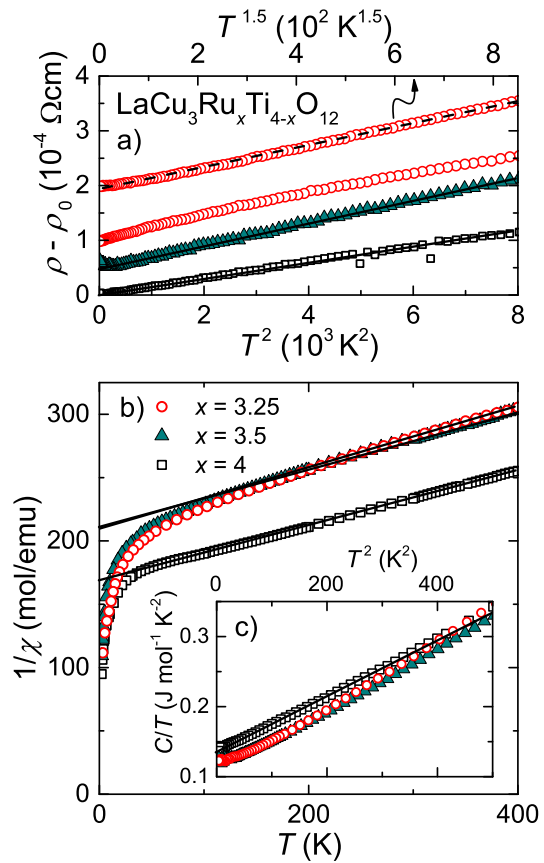}%
\caption{(Color online) Transport, magnetic and thermodynamic properties of LaCu$_3$Ru$_x$Ti$_{4-x}$O$_{12}$ for $x > 3$. a) Resistivity after subtraction of its zero-temperature value dependent on squared temperature $T^2$ (lower abscissa) and in addition for $x=3.25$ as function of $T^{1.5}$ (upper abscissa). The curves are shifted vertically to each other for clarity. b) Inverse magnetic susceptibility $1/\chi$ as a function of temperature. c) Specific heat in representation $C/T$ vs. $T^2$. All straight lines mark the linear ranges of the measured data in the respective representations. \label{Ru-High}}
\end{figure}

To study the evolution of the resistivity on ruthenium dilution in more detail, Fig.~\ref{Ru-High} compares the temperature dependence of resistivity (a), susceptibility (b), and specific heat (c) for samples with high Ru content $x > 3$. Starting with the resistivity, we observe the quadratic temperature dependence typical for heavy-fermion metals for all three samples under consideration.
To corroborate the heavy-fermion behavior, we focus on the inverse magnetic susceptibility $1/\chi(T)$ and low-temperature specific heat in the representation $C/T$ vs. $T^2$. For $T > 100$~K, the susceptibility of all three compounds is well described by an antiferromagnetic Curie-Weiss law with a Curie-Weiss temperature $\Theta_{\rm CW}$ of the order of $-10^3$~K and an effective paramagnetic moment $\mu_{\rm eff}$ compatible with the sum of the corresponding number of Cu$^{2+}$ ($S=1/2$) and Ru$^{4+}$ ($S=1$) spins per formula unit (cf. Tab~\ref{Tab:Wilson}). In contrast, the Curie-Weiss tail at low temperatures is due to small amounts of weakly coupled spins. Already the fact that there is no magnetic order down to lowest temperatures is in favor of a Fermi-liquid state. In classical $f$-based heavy-fermion compounds the empirical relation between Curie-Weiss temperature and the characteristic temperature is given by $T^* \sim \Theta_{\rm CW}/4$ (Ref. \onlinecite{Maple96}). For $x = 4$, this rule of thumb yields $T^* = 190$\,K in agreement with the kink in the resistivity data (cf. Fig. \ref{Ru_x3and4}), once more supporting the Kondo-like physics here.
From the extrapolation of the extremely flat -- almost Pauli-like -- Curie-Weiss law to $T=0$ we determined the zero-temperature susceptibility $\chi(0)$. Fitting the low-temperature specific heat by $C=\gamma T + \beta T^3$ the Sommerfeld coefficient $\gamma$ as well as the phonon contribution $\beta$ were obtained. Both, $\chi(0)$ and $\gamma$ are significantly enhanced when compared to common metals. Tables~\ref{Tab:Wilson} and \ref{Tab:Kadowaki-Woods} list the corresponding parameters and the resulting Wilson ratio $R_{\rm W}= [4 \pi^2 k_{\rm B}^2] / [3 g^2 \mu_{\rm B}^2] \times \chi(0)/\gamma$ (Ref. \onlinecite{Wilson75}) and Kadowaki-Woods ratio $R_{\rm KW}=A/\gamma^2$ (Ref. \onlinecite{Kadowaki86}). The values of both ratios fit well into the typical range found for heavy-fermion compounds, thus confirming the Fermi-liquid properties for $x>3$ comparable to canonical heavy-fermion systems close to a quantum-critical point like CeCu$_6$ or CeRu$_2$Si$_2$ (cf. Fig.~1 in Ref.~\onlinecite{Krug2003}).

\begin{table}
\centering
\caption{Curie-Weiss temperature $\Theta_{\rm CW}$, effective paramagnetic moment $\mu_{\rm eff}$, theoretical effective moment $\mu_{\rm th}$, extrapolated zero-temperature susceptibility $\chi(0)$, Sommerfeld coefficient $\gamma$, and corresponding Wilson ratio $R_\mathrm{W}$ of LaCu$_3$Ru$_x$Ti$_{4-x}$O$_{12}$ for $x\geq3.25$.\label{Tab:Wilson}}
\begin{tabular}{ccccccc}
\noalign{\smallskip}\hline \hline \noalign{\smallskip}
 $x$ & $\Theta_{\rm CW} (\text{K})$ & $\mu_{\rm eff} (\mu_{\rm B})$ & $\mu_{\rm th} (\mu_{\rm B})$ & $\chi(0) (\frac{\text{memu}}{\text{mol}})$ & $\gamma (\frac{10^3 \text{erg}}{\text{mol}\,\text{K}^2})$ & $R_\mathrm{W}$ \\
\noalign{\smallskip}\hline \noalign{\smallskip}
    3.25 & -909 & 5.85 & 5.76 & 4.71 & 1135 & 3.03 \\
    3.5 & -884 & 5.81 & 5.91 & 4.76 & 1154 & 3.01 \\
    4 & -762 & 6.03 & 6.21 & 5.98  & 1376 & 3.17 \\
\noalign{\smallskip}\hline \hline
\end{tabular}
\end{table}

\begin{table}
\centering
\caption{Low-temperature resistivity coefficient $A$, Sommerfeld coefficient $\gamma$ normalized by the Ru concentration $x$, and resulting Kadowaki-Woods ratio per Ru ion $R_\mathrm{KW}$ of LaCu$_3$Ru$_x$Ti$_{4-x}$O$_{12}$ for $x\geq3.25$.\label{Tab:Kadowaki-Woods}}
\begin{tabular}{cccc}
\noalign{\smallskip}\hline \hline \noalign{\smallskip}
 $x$ & $A (\frac{10^{-8}\Omega\,\text{cm}}{\text{K}^2})$ & $\frac{\gamma}{x} (\frac{10^{-3}\text{J}}{\text{mol}\,\text{K}^2})$ & $R_\mathrm{KW} (\frac{10^{-5}\Omega\,\text{cm}\,\text{mol}^2\,\text{K}^2}{\text{J}^2})$ \\
\noalign{\smallskip}\hline \noalign{\smallskip}
    3.25 & 2.66 & 34.9 & 2.18 \\
	3.5 & 2.08 & 33.0 & 1.91 \\
	4 & 1.54 & 34.4 & 1.30 \\
\noalign{\smallskip}\hline \hline
\end{tabular}
\end{table}

\begin{figure}
\includegraphics[width=0.4\textwidth]{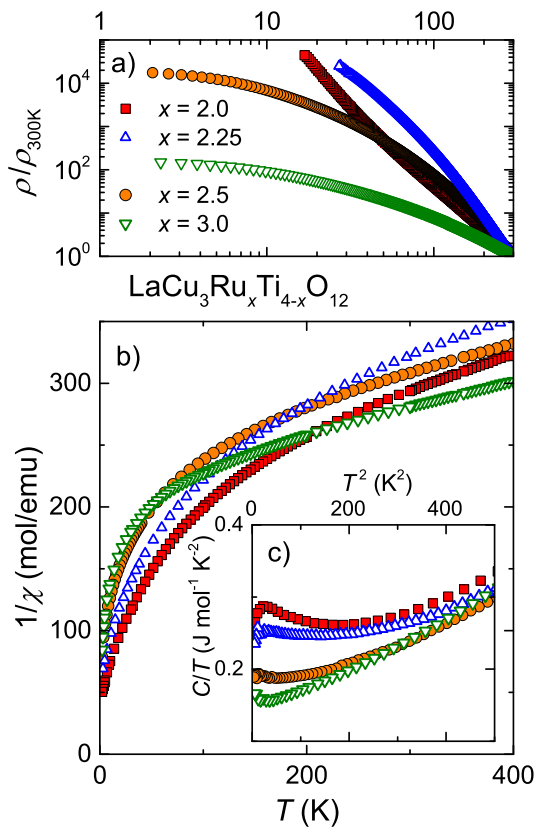}%
\caption{(Color online) Transport, magnetic and thermodynamic properties of LaCu$_3$Ru$_x$Ti$_{4-x}$O$_{12}$ for $2 \leq x \leq 3$. a) Resistivity normalized to its room-temperature value at 300~K on logarithmic scales. b) Inverse magnetic susceptibility $1/\chi$ as a function of temperature. c) Specific heat in representation $C/T$ vs. $T^2$.\label{Ru-Mid}}
\end{figure}

It has to be noted that on approaching intermediate Ru concentrations the resistivity of $x=3.25$ is even better described by a $\rho \propto T^{3/2}$ temperature-dependent power law suggesting a non-Fermi-liquid behavior close to a quantum-critical point at the transition to a disorder-dominated antiferromagnet.\cite{Moriya95,Sengupta95,vLoeh2007} Moreover, as shown in Fig.~\ref{Ru-Mid}(a) for further dilution $x \leq 3$ the metallic characteristics breaks down and the resistivity increases on decreasing temperature. Indeed, on decreasing temperature not only for $x=3$ but also for $x=2.5$ and 2.25, the resistivity $\rho(T)$ is approaching saturation towards lowest temperatures, as expected for the single-ion Kondo effect.\cite{Kond64} Note, however, that in contrast to those $4f$ systems, in LaCu$_3$Ru$_x$Ti$_{4-x}$O$_{12}$ the absolute increase of the resistivity to low temperatures is much stronger, because on dilution of ruthenium by titanium the conduction electrons tend to increasingly localize at the copper sites. This localization is corroborated by combined results of x-ray absorption near-edge structure (XANES) measurements and thermogravimetric determination of the oxygen stoichiometry.\cite{Riegg15} While the Ru-K absorption edge and, hence, the Ru valence remains almost unchanged at $+4$, the copper valence increases linearly from $+1.6$ $(x = 4)$ to $+2$ $(x = 0)$ with decreasing ruthenium concentration. Furthermore, the change of the shape of the Cu $K$-absorption edge reflects the variation of the localization of the Cu $3d$ electrons. For $x < 2$ rather sharp peak-like structures are observable, which become visibly smoothed in the ruthenium-rich metallic regime. From this an increasing delocalized character of the Cu electrons can be derived, while the Ru $4d$ electrons appear to be localized over the whole concentration range. Additionally, in the spin-glass regime $0.5 \leq x \leq 1$ a significant oxygen excess is observed, which stabilizes the obtained progress of Ru and Cu oxidation states.

Therefore, due to the localization of conduction electrons at the copper site the conductivity gradually decreases and, in turn, the single-ion Kondo-like behavior transforms into a purely semiconducting characteristics for $x$ approaching 2. Regarding the susceptibility (b) for $x<3$ in Fig.~\ref{Ru-Mid}, there is no simple Curie-Weiss law visible in $1/\chi(T)$, anymore. Moreover, an increasing excess contribution shows up in the specific heat $C/T$ vs. $T^2$ (cf. Fig. \ref{Ru-Mid}(c)). Both is related to the dilution of the Ru lattice and resulting distribution of Curie-Weiss temperatures $\Theta_{\rm CW}$ and corresponding $T^*$.

\begin{figure}[b]
\includegraphics[width=0.45\textwidth]{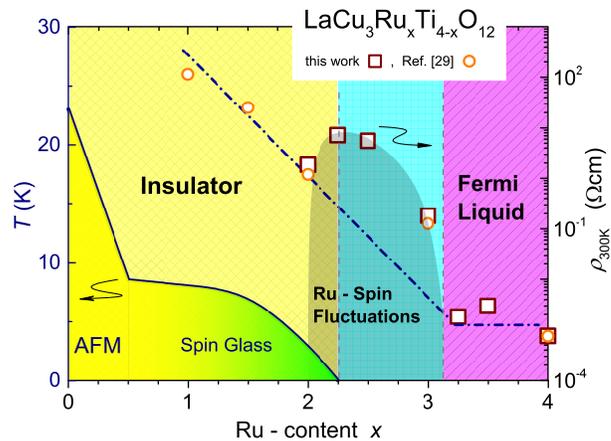}%
\caption{(Color online) Schematic phase diagram of LaCu$_3$Ru$_x$Ti$_{4-x}$O$_{12}$ based on the the room-temperature resistivity $\rho_{\rm 300K}$ as a function of the ruthenium content $x$ (right ordinate). The temperatures of magnetic transitions into antiferromagnetic (AFM) and spin-glass states, respectively, are taken from Ref. \onlinecite{Riegg15a} (left ordinate). Lines are drawn to guide the eye. \label{phasediagram}}
\end{figure}

To summarize the transport properties, Fig.~\ref{phasediagram} illustrates a schematic phase diagram based on the dependence of the absolute resistivity at room temperature $\rho_{\rm 300K}$ as function of the Ru content $x$ (right ordinate): for $x > 3$ the value of $\rho_{\rm 300K}$ remains nearly independent on $x$ of the order of $10^{-3}$\,$\Omega$cm. On further decreasing ruthenium content one observes an increase by two orders of magnitude at $x=3$ followed by a dome with a maximum $\rho_{\rm 300K} \sim 10$\,$\Omega$cm at $x=2.25$ and a decrease by one order of magnitude at $x=2$. In the insulating regime for $x \leq 2$ the room-temperature resistivity finally increases monotonously to lower $x$ (\textit{cf.} dashed line in Fig.~\ref{phasediagram}). One clearly recognizes the enhanced resistivity values in the metallic regime where the dilution of the ruthenium lattice breaks the coherent Fermi-liquid state and obviously the Ru-spin fluctuations provoke an additional contribution to the electric resistivity.

\section{Magnetic Resonance}

\begin{figure}
\includegraphics[width=0.4\textwidth]{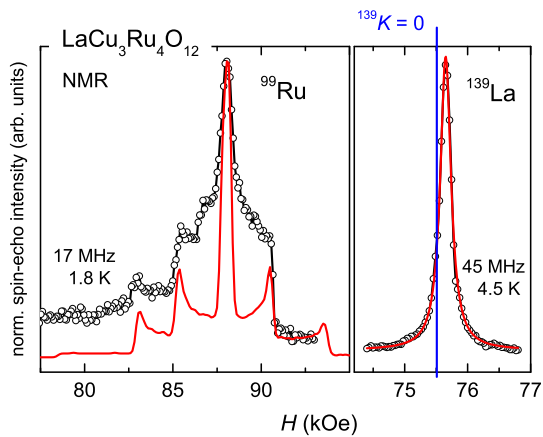}
\caption{(Color online) Low-temperature NMR spectra of $^{99}$Ru and $^{139}$La. The solid red line simulates the electric quadrupole interaction in case of $^{99}$Ru as described for CaCu$_3$Ru$_4$O$_{12}$ in Ref.~\onlinecite{Kri08}. For $^{139}$La, the solid red line is a fit to a Lorentzian line which exhibits a negative line shift from the diamagnetic reference position $^{139}K = 0$ in LaCl$_3$ (blue line). \label{NMR_spectra}}
\end{figure}

\begin{figure}
\includegraphics[width=0.45\textwidth]{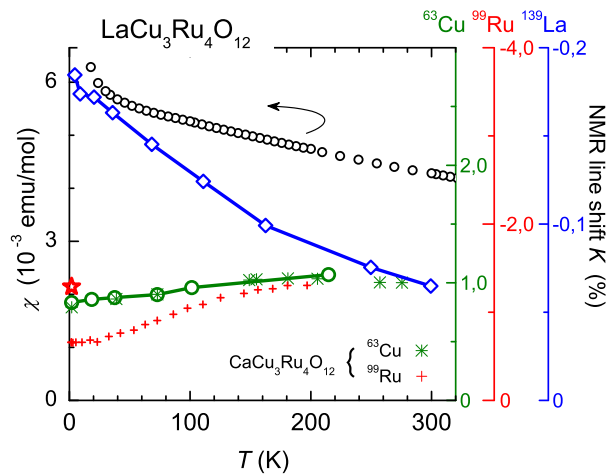}
\caption{(Color online) Bulk susceptibility $\chi(T)$ (open black circles, left scale) and local susceptibility represented by NMR line shift $K(T)$ (open symbols, right scales) of LaCu$_3$Ru$_4$O$_{12}$. Line shifts of $^{63}$Cu (green spheres/right scale) and $^{139}$La
(open blue diamonds/rightmost scale) are given, respectively. In addition one single low-temperature value of $^{99}$Ru line shift $K = - 1.29$\% is plotted at $T = 1.8$~K (red asterisk). For comparison corresponding data of CaCu$_3$Ru$_4$O$_{12}$ is plotted as solid symbols (taken from Ref. \onlinecite{Kri09}).   \label{NMR_line_shifts}}
\end{figure}

For a closer inspection of the Curie-Weiss laws $\chi(T)$ with extremely large Curie-Weiss temperatures displayed in Fig. \ref{Ru-High} we employed nuclear magnetic resonance (NMR) for LaCu$_3$Ru$_4$O$_{12}$ and were able to elucidate the situation from the local susceptibility point of view: NMR spectra of the isotopes $^{139}$La, $^{99}$Ru, and $^{63}$Cu were obtained as exemplarily documented in Fig.~\ref{NMR_spectra} for the two former isotopes. The latter one, $^{63}$Cu, exhibits similar spectra as already shown for the calcium homologue compound CaCu$_3$Ru$_4$O$_{12}$ in Ref.~\onlinecite{Kri08} with slightly smaller values of the local electric field gradient here. While $^{99}$Ru with a nuclear spin $I=5/2$ exhibits a distinct quadrupole splitting,\cite{Kri08} the spectrum of $^{139}$La ($I=7/2$) can be well described in terms of a single Lorentz line.

The temperature dependences of the line shift $K(T)$ of all isotopes are shown in Fig.~\ref{NMR_line_shifts} (right ordinates) together with bulk susceptibility measurements (left ordinate). Towards decreasing temperatures $T < 200$~K, the local susceptibility $K(T)$ at the copper site experiences the same decrease due to Kondo-like moment compensation as we reported already for the electronic correlations active in CaCu$_3$Ru$_4$O$_{12}$ (\textit{cf.} Ref.~\onlinecite{Kri09}). On the other hand, the line shift $K(T)$ at the lanthanum site exhibits a Curie-like increase towards lower temperatures which documents transferred hyperfine fields from fluctuating paramagnetic ruthenium moments. Due to these strong spin fluctuations of the ruthenium magnetic moment we observed a weak $^{99}$Ru nuclear signal only and succeeded to collect an NMR spectrum only at the lowest temperature shown in Figure \ref{NMR_spectra}. This spectrum nevertheless exhibits a line shift value $K({}^{99}{\rm Ru}) = - 1.29$~\% which strongly exceeds the value of $-0.71$~\% observed at low temperatures $T < 40$\,K in the case of CaCu$_3$Ru$_4$O$_{12}$ where a strong moment compensation was detected at the ruthenium site. Note that in CaCu$_3$Ru$_4$O$_{12}$ a $^{99}$Ru-line shift comparable to LaCu$_3$Ru$_4$O$_{12}$ is reached only for elevated temperatures $T \sim 150$\,K. Recalling the valence instability reported for CaCu$_3$Ru$_4$O$_{12}$ at $T = 150$\,K, the absence of such a strongly reduced $^{99}$Ru-line shift proves the stability of the Ru$^{4+}$ tetravalent state in LaCu$_3$Ru$_4$O$_{12}$ down to lowest temperatures. Therefore, only the ruthenium magnetic moments in LaCu$_3$Ru$_4$O$_{12}$ are responsible for the Curie-Weiss-type susceptibility, whereas the copper magnetic moments appear to be fully compensated by conduction electrons. Nevertheless, moment compensation is important at the ruthenium site as well, in order to explain the deviation of the $^{139}$La Knight shift from the pure Curie-Weiss law in favor of a temperature independent susceptibility below the coherence temperature $T_{\rm coh} \sim 50$\,K, an unambiguous fingerprint of the Ru$^{4+}$ Kondo lattice in LaCu$_3$Ru$_4$O$_{12}$.

\begin{figure}
\includegraphics[width=0.45\textwidth]{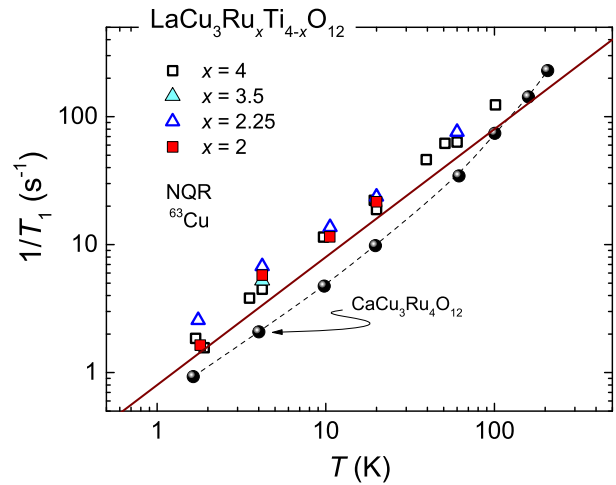}
\caption{(Color online) Temperature dependence of the NQR spin-lattice relaxation rate $1/T_1$ for LaCu$_3$Ru$_x$Ti$_{4-x}$O$_{12}$ compared to the data of CaCu$_3$Ru$_4$O$_{12}$ taken from Ref.~\onlinecite{Kri09}. The solid line is the Korringa linear increase $T_1T=1.26$~sK of pure copper metal taken from Ref.~\onlinecite{Han72}. \label{NMR_T1}}
\end{figure}

\begin{figure}
\includegraphics[width=0.4\textwidth]{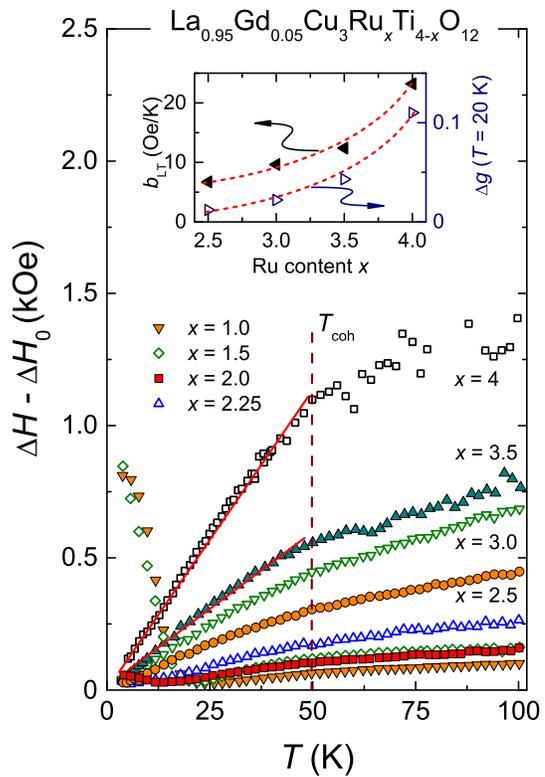}%
\caption{(Color online) Temperature dependence of the ESR linewidth $\Delta H - \Delta H_0$ of 5\% Gd substituted LaCu$_3$Ru$_x$Ti$_{4-x}$O$_{12}$ after subtraction of the temperature independent residual contribution $\Delta H_0$. Solid lines mark the Korringa-like linear increase. Inset: low-temperature Korringa slope $b_{\rm LT}$ and the $g$ shift from its typical value in insulators dependent on the ruthenium content. The dotted lines indicate the simultaneous behavior of both quantities. \label{ESR}}
\end{figure}

The local electronic properties on dilution of the ruthenium lattice were further analyzed by resonance spectroscopic methods NQR and ESR. The corresponding results are shown in Figs.~\ref{NMR_T1} and \ref{ESR}, respectively, focussing on the longitudinal spin-relaxation rates $1/T_1$ obtained by both methods. In NQR the relaxation rate $1/T_1$ is directly measured by pulse methods, whereas in ESR it is extracted from the half-width-at-half-maximum (HWHM) line width of the exchange-narrowed Lorentz-shaped Gd resonance. A temperature-independent residual line width $\Delta H_0$ has to be subtracted such that the extrapolated difference $\Delta H - \Delta H_0$ vanishes at $T=0$ (Ref. \onlinecite{Schmi14}). $\Delta H_0$ results from inhomogeneities and impurities. The difference $\Delta H - \Delta H_0$ resembles the transverse relaxation rate $1/T_2$, but $T_1=T_2$ usually holds in metals.\cite{Bar81} In conventional metals $1/T_1$ is expected\cite{Kor50} to follow a linear Korringa law $1/T_1 = b \cdot T$, where the slope probes the squared electronic density of states at the Fermi level $b \propto N^2(E_{\rm F}) \propto (m^*)^2$. Indeed, as illustrated in Fig.~\ref{NMR_T1}, in the whole concentration range $2 \leq x \leq 4$ the relaxation rate of the $^{63}$Cu-NQR exhibits a linear temperature dependence with slightly enhanced slope as compared to copper metal.\cite{Han72} This is in fair agreement with the observation in CaCu$_3$Ru$_4$O$_{12}$ (Ref. \onlinecite{Kri09}), where $1/T_1$ of $^{63}$Cu is even slightly lower than in metallic copper due to the valence fluctuations. This result shows that locally at the copper site, LaCu$_3$Ru$_x$Ti$_{4-x}$O$_{12}$ exhibits nearly conventional metallic characteristics in the whole Ru rich range $2 \leq x \leq 4$ with only weak enhancement of $N(E_{\rm F})$. This astonishing concentration-independent behavior indicates that the Cu nuclei appear to be decoupled from the magnetic Ru sublattice in strong conflict with conventional intermetallic behavior. In intermetallic systems one expects the existence of the Ruderman-Kittel-Kasuya-Yosida (RKKY) interaction isotropically mediated by conduction electrons between localized moments, i.e. also between electronic and nuclear local moments. Note, however that in oxides super-exchange interactions rather than the RKKY interaction have to be considered due to the anisotropy of the orbital geometries coupling the local moments. Therefore, the local electronic density of states $N^2(E_{\rm F})$ at the Cu site is hardly influenced by the Ru concentration.

A different situation appears from the viewpoint of Gd-ESR: for $x=4$ the line width of the Gd resonance exhibits a significantly enhanced linear increase $b_{\rm LT} \approx 24$~Oe/K at temperatures below 50~K as compared to conventional metals with a Korringa slope of about $b \approx 5$~Oe/K (Refs. \onlinecite{Krug2003,Schl88}). Above 50~K the line-width data $\Delta H(T)$ deviate from linearity into a weaker monotonous increase. This is just the behavior expected for heavy fermions, a strong linear increase for low temperatures $T \leq T_{\rm coh} \ll T^*$ followed by a nonlinear regime which finally approaches a moderate Korringa law at high temperatures.\cite{Col87} The coherence temperature $T_{\rm coh}$ denotes the upper temperature limit of the Fermi-liquid regime.\cite{Ott87,Grewe91} Thus, Gd-ESR is sensitive to the electronic mass enhancement at the ruthenium ions, which was directly measured by $^{101}$Ru-NQR in CaCu$_3$Ru$_4$O$_{12}$ (Ref. \onlinecite{Kri09}). However, the mass enhancement probed by the Gd-ESR turns out to be significantly smaller than that measured by $^{101}$Ru-NQR. This corroborates the picture of a local Kondo effect at the ruthenium site and can be understood in terms of the RKKY-like transfer of the Ru spin fluctuations to the Gd spin via the conduction electrons.\cite{Col87} The Ru spin fluctuations just result in an additional relaxation contribution $1/T_1 \propto T \chi_{\rm Ru} \tau_{\rm Ru}$, where $\chi_{\rm Ru}$ and $\tau_{\rm Ru}$ denote the susceptibility of the Ru spins and their fluctuation time, respectively. For $T \leq T_{\rm coh}$ both are practically temperature independent, resulting in an enhanced Korringa-like contribution to the relaxation rate. On substitution of ruthenium by titanium the qualitative temperature dependence of the gadolinium line width remains comparable with a similar $T_{\rm coh}$. However, a gradually decreasing slope $b_{\rm LT}$ down to 7.5~Oe/K for $x=2.5$ is obtained due to the dilution of the ruthenium lattice. Finally for $x \leq 2$, the Korringa regime vanishes and the line width even increases on approaching the spin-glass state.

The dependence of the $g$ value on the Ru content $x$ completes the picture: the $g$ shift ($\Delta g = 1.993 - g$) from its value typically found in insulators \cite{Abr70} locally measures the susceptibility at the Gd site, which mainly results from the magnetic polarization of the neighboring Ru spins, i.e. $\Delta g \propto \chi_{\rm Ru}$. Hence, the inset in Fig.~\ref{ESR} compares the low-temperature Korringa-like slope $b_{\rm LT}$ with the $g$ shift found below $T_{\rm coh}$. Indeed, both exhibit a very similar behavior once more supporting the Kondo-effect at the ruthenium site to be responsible for the heavy-fermion properties of LaCu$_3$Ru$_x$Ti$_{4-x}$O$_{12}$. One could infer that Moriya's spin-fluctuation theory \cite{Moriya1979} accounts for the electronic mass enhancement in LaCu$_3$Ru$_4$O$_{12}$. In that case one would expect a divergence of the spin-lattice relaxation rate close to the magnetic instability,\cite{Ishigaki96,Hei01,Krug2003} which in LaCu$_3$Ru$_4$O$_{12}$ coincides with the MIT. However, such a divergent behavior is not observed in the magnetic resonance experiments presented here. Therefore, the Kondo-effect appears to be the most applicable scenario in these ruthenates under investigation.

\section{Conclusion}
To conclude, we have compared magnetic, thermodynamic, and transport properties of LaCu$_3$Ru$_x$Ti$_{4-x}$O$_{12}$ with $^{63}$Cu, $^{99}$Ru, and $^{139}$La NMR/NQR as well as Gd$^{3+}$ ESR to analyze the heavy-fermion behavior at the Ru-rich side of the phase diagram above the metal-to-insulator transition $x \geq 2.25$. The quadratic low-temperature behavior of the electrical resistivity, the enhanced Sommerfeld coefficient of the specific heat, and the large Pauli-like low-temperature susceptibility characterize the compounds with high ruthenium content $x \geq 3.25$ as heavy-fermions with moderately enhanced effective masses. The corresponding Wilson and Kadowaki-Woods ratios are comparable to those of canonical $4f$-based heavy-fermion compounds. In the adjacent regime of intermediate ruthenium content $2.25 \leq x \leq 3$ the coherent heavy-fermion state breaks down giving rise to a Kondo-like increase of the resistivity on decreasing temperature as typically observed in canonical $4f$-based Kondo lattices on dilution of the Kondo ions. This observation suggests the spin fluctuations of the $4d$ Ru$^{4+}$ ions due to their screening by the conduction electrons to be responsible for the heavy-fermion formation.

This hypothesis is corroborated by the NMR and ESR measurements probing the local susceptibility at different lattice sites: while valence fluctuations strongly reduce the $^{99}$Ru-NMR line shift in CaCu$_3$Ru$_4$O$_{12}$, the unreduced shift proves
the stability of the ruthenium valence in LaCu$_3$Ru$_4$O$_{12}$ down to lowest temperatures. At the same time the $^{63}$Cu line shift is comparable in both compounds. However, the line shift at the lanthanum site in LaCu$_3$Ru$_4$O$_{12}$ reproduces the Curie-Weiss behavior of the static susceptibility with a tendency for saturation below 50~K in agreement with the coherent Fermi-liquid state at $T<T_{\rm coh}$. From the NQR spin-lattice relaxation rates at the different lattice sites analogous information is obtained: while the $^{63}$Cu $1/T_1$ is not significantly enhanced with respect to normal metals, the $^{101}$Ru $1/T_1$ appears to be enormously enhanced - in the La compound even stronger than in the Ca compound - as Ru NMR spectra could be only obtained at lowest temperature in LaCu$_3$Ru$_4$O$_{12}$ due to fast relaxation. Substitution of ruthenium by titanium does not affect the slow $^{63}$Cu $1/T_1$ spin-lattice relaxation rate indicating the copper site to be nearly decoupled from the processes at the ruthenium site. Finally, Gd-ESR at the lanthanum site yields the spin-lattice relaxation rate Gd$^{3+}$ $1/T_1$ typical for a heavy-fermion system with $T_{\rm coh}=50$\,K proportional to the local susceptibility. On substitution of ruthenium by titanium this relaxation contribution is gradually diminished, once more proving the ruthenium ions to be responsible for heavy-fermion formation.

These combined experiments allow identifying the Kondo effect and resulting spin flucuations of the localized Ru$^{4+}$ spins to be the source of heavy-fermion formation in LaCu$_3$Ru$_4$O$_{12}$. The localized nature of the ruthenium spins is supported by the local properties proven by nuclear magnetic resonance and electron spin resonance. This resembles the very similar behavior observed in $4f$ based intermetallic Kondo-lattice compounds. Frustration as a source of heavy-fermion behavior can be ruled out, since the Ru sites form a simple cubic sublattice. Dilution of the Ru$^{4+}$ Kondo lattice by nonmagnetic Ti$^{4+}$ results in single-ion Kondo properties, which gradually weaken with increasing Ti concentration and finally are suppressed at the metal-to-insulator transition, where the itinerant electrons localize at the Cu sites. In addition, this localization gives rise to the quantum phase transition into the spin-glass state without any mass enhancement, ruling out the metal-to-insulator transition as source of heavy-fermion formation.

\begin{acknowledgments}
We gratefully acknowledge financial support granted by the Bavarian graduate school (Resource strategy concepts for sustainable energy systems) of the Institute of Materials Resource Management (MRM) of the University of Augsburg and by the Deutsche Forschungs Gemeinschaft (DFG) within the collaborative research center TRR~80 (Augsburg, Munich). A.~G. was supported in the frame of the DFG research unit FOR~960.
\end{acknowledgments}


\begin{thebibliography}{}

\bibitem{Kond97}
S. Kondo, D. C. Johnston, C. A. Swenson, F. Borsa, A. V. Mahajan, L. L. Miller, T. Gu, A. I. Goldman, M. B. Maple, D. A. Gajewski, E. J. Freeman, N. R. Dilley, R. P. Dickey, J. Merrin, K. Kojima, G. M. Luke, Y. J. Uemura, O. Chmaissem, and J. D. Jorgensen, Phys. Rev. Lett. \textbf{78}, 3729 (1997).

\bibitem{Kri99}
A. Krimmel, A. Loidl, M. Klemm, S. Horn, and H. Schober,
Phys. Rev. Lett. \textbf{82}, 2919 (1999).

\bibitem{And75}
K. Andres, J. E. Graebner, and H. R. Ott, Phys. Rev. Lett. \textbf{35}, 1779 (1975).

\bibitem{Ste79}
F. Steglich, J. Aarts, C. D. Bredl, W. Lieke, D. Meschede, W. Franz, and H. Sch\"afer, Phys. Rev. Lett. \textbf{43}, 1892 (1979).

\bibitem{Tro00}
O. Trovarelli, C. Geibel, S. Mederle, C. Langhammer, F. M. Grosche, P. Gegenwart, M. Lang, G. Sparn, and F. Steglich, Phys. Rev. Lett. \textbf{85}, 626 (2000).

\bibitem{Ste84}
G. R. Stewart, Rev. Mod. Phys. \textbf{56}, 755 (1984).

\bibitem{Kond64}
J. Kondo, Prog. Theoret. Phys. \textbf{32}, 505 (1964).

\bibitem{Fulde01}
P. Fulde, A. N. Yaresko, A. A. Zvyagin, and Y. Grin,
Europhys. Lett. \textbf{54}, 779 (2001).


\bibitem{Tok93}
Y. Tokura, Y. Taguchi, Y. Okada, Y. Fujishima, T. Arima, K. Kumagai, and Y. Iye, Phys. Rev. Lett. \textbf{70}, 2126 (1993).

\bibitem{Tag93}
Y. Taguchi, Y. Tokura, T. Arima, and F. Inaba, Phys. Rev. B \textbf{48}, 511 (1993).

\bibitem{Hei01}
M. Heinrich, H.-A. Krug von Nidda, V. Fritsch, and A. Loidl, Phys. Rev. B \textbf{63}, 193103 (2001).

\bibitem{Cao97} G. Cao, S. McCall, M. Shepard, J. E. Crow, and R. P. Guertin, Phys. Rev. B \textbf{56}, 321 (1997).

\bibitem{Mae94} Y. Maeno, H. Hashimoto, K. Yoshida, S. Nishizaki, T.
Fujita, J.G. Bednorz, and F. Lichtenberg, Nature \textbf{372}, 532
(1994).

\bibitem{Mae01} Y. Maeno, T.M. Rice, and M. Sigrist, Phys. Today \textbf{54},
42 (2001).

\bibitem{Gri01} S. A. Grigera, R. S. Perry, A. J. Schofield, M. Chiao, S. R. Julian, G. G. Lonzarich, S. I. Ikeda, Y. Maeno,
A. J. Millis, and A. P. Mackenzie, Science \textbf{294}, 329 (2001).

\bibitem{Kha01} P. Khalifah, K. D. Nelson, R. Jin, Z. Q. Mao, Y. Liu, Q. Huang, X. P. A. Gao, A. P. Ramirez, and R. J. Cava,
Nature \textbf{411}, 669 (2001).

\bibitem{Tan09}
S. Tanaka, N. Shimazui, H. Takatsu, S. Yonezawa, and Y. Maeno, J. Phys. Soc. Jpn. \textbf{78}, 024706 (2009).

\bibitem{Krug2003} H.-A. Krug von Nidda, R. Bulla, N. B\"uttgen, M. Heinrich, and A. Loidl, Eur. Phys. J. B \textbf{34}, 399 (2003).

\bibitem{Ebb02}
S. G. Ebbinghaus, A. Weidenkaff, and R. J. Cava, J. Solid State Chem. \textbf{167}, 126 (2002).

\bibitem{Hebert15}
S. Hebert, R. Daou, and A. Maignan, Phys. Rev. B \textbf{91}, 045106 (2015).

\bibitem{Kob04}
W. Kobayashi, I. Terasaki, J. Takeya, I. Tsukada, and Y. Ando, J. Phys. Soc. Japan \textbf{73}, 2373 (2004).

\bibitem{Kri08}
A. Krimmel, A. G\"unther, W. Kraetschmer, H. Dekinger, N. B\"uttgen, A. Loidl, S. G. Ebbinghaus, E. W. Scheidt, W. Scherer, Phys. Rev. B \textbf{78}, 165126 (2008).

\bibitem{Kri09}
A. Krimmel, A. G\"unther, W. Kraetschmer, H. Dekinger, N. B\"uttgen, V. Eyert, A. Loidl, D. V. Sheptyakov, E.-W. Scheidt, and W. Scherer, Phys. Rev. B \textbf{80}, 121101 (2009).

\bibitem{Hom01} C. C. Homes, T. Vogt, S. M. Shapiro, S. Wakimoto, and A. P. Ramirez, Science \textbf{293}, 673 (2001).

\bibitem{Lun04} P. Lunkenheimer, R. Fichtl, S. G. Ebbinghaus, and A. Loidl, Phys. Rev. B \textbf{70}, 172102 (2004).

\bibitem{Ram04}
A. P. Ramirez, G. Lawes, D. Li, and M. A. Subramanian, Solid State Commun. \textbf{131}, 251 (2004).

\bibitem{Tsu09}
I. Tsukada, R. Kammuri, T. Kida, S. Yoshii, T. Takeuchi, M. Hagiwara, M. Iwakawa, W. Kobayashi, and I. Terasaki, Phys. Rev. B \textbf{79}, 054430 (2009).

\bibitem{Buet10}
N. B\"uttgen, H.-A. Krug von Nidda, W. Kraetschmer, A. G\"unther, S. Widmann, S. Riegg, A. Krimmel, and A. Loidl, J. Low Temp. Phys. \textbf{161}, 146 (2010).


\bibitem{Ebb10} S. G. Ebbinghaus, S. Riegg, T. G\"otzfried, and A. Reller, Eur. Phys. J. ST \textbf{180}, 91 (2010).

\bibitem{Schmi14} B. Schmidt, H.-A. Krug von Nidda, S. Riegg, S. G. Ebbinghaus, A. Reller, and A. Loidl,
Magn. Reson. Solids \textbf{16}, 14210 (2014).

\bibitem{Riegg15a} S. Riegg, S. Widmann, A. G\"unther, B. Meir, S. Wehrmeister, S.Sterz, W. Kraetschmer, S.G. Ebbinghaus, A. Reller, N. B\"uttgen, H.-A. Krug von Nidda, and A. Loidl, Eur. Phys. J. Special Topics \textbf{224}, 1061 (2015).

\bibitem{Riegg15}
S. Riegg, A. Reller, A. Loidl, and S. G. Ebbinghaus, Dalton Transactions {\bf 44}, 10852 (2015).

\bibitem{rho} For these experiments the polycrystalline material was pelletized. In case of the resistivity measurements square bars of $13 \times 2 \times 2$~mm$^3$ size were prepared. All pellets were sintered at 1000$^{\circ}$C for 10 hours and contacts were made up of silver paint in a four-point setup.

\bibitem{NQR} The homodyne demodulation of the radio-frequency (RF) nuclear signals was achieved by digital quadrature detection using a fast two-channel oscilloscope card.

\bibitem{ESR} To reduce the skin effect due to conductivity the ESR and NQR measurements were performed on fine powder.



\bibitem{Buet97}
N. B\"uttgen, H. -A. Krug von Nidda, and A. Loidl
Physica B \textbf{230-232}, 590 (1997).

\bibitem{KvN98}
H.-A. Krug von Nidda, A. Sch\"utz, M. Heil, B. Elschner, and A. Loidl,
Phys. Rev. B \textbf{57}, 14344 (1998).

\bibitem{Sereni80} J. G. Sereni, J. Phys. F: Metal Phys. \textbf{10} 2831 (1980).

\bibitem{Umehara90} I. Umehara, Y. Kurosawa, N. Nagai, M. Kikuchi, K. Satoh, Y. Onuki, J. Phys. Soc. Jpn. \textbf{59}, 2848 (2009).

\bibitem{Ragel04} F. C. Ragel, P. V. du Plessis, J. Phys.: Condens. Matter \textbf{16} 2647 (2004).

\bibitem{Ock01}
M. Ocko, D. Drobac, B. Buschinger, C. Geibel, and F. Steglich, Phys. Rev. B \textbf{64}, 195106 (2001).

\bibitem{Maple96} M. B. Maple, R. P. Dickey, J. Herrmann, M. C. de Andrade, E. J. Freeman, D. A.
Gajewski, and R. Chau, J. Phys.: Condens. Matter \textbf{8}, 9773 (1996).

\bibitem{Wilson75}
K.G. Wilson, Rev. Mod. Phys. {\bf 47}, 773 (1975).

\bibitem{Kadowaki86}
K. Kadowaki, S.B. Woods, Solid State Commun.
{\bf 58}, 507 (1986).

\bibitem{Moriya95}
T. Moriya and T. Takimoto, J. Phys. Soc. Jpn. \textbf{64}, 960 (1995).

\bibitem{Sengupta95}
A. M. Sengupta and A. Georges, Phys. Rev. B \textbf{52}, 10295 (1995).

\bibitem{vLoeh2007}
H. v. L\"ohneysen, A. Rosch, M. Vojta, and P. W\"olfle, Rev. Mod. Phys. \textbf{79}, 1015 (2007).

\bibitem{Bar81}
S. E. Barnes, Adv. Phys. \textbf{30}, 801 (1981).

\bibitem{Kor50}
J. Korringa, Physica \textbf{16}, 601 (1950).

\bibitem{Han72}
M. Hanabusa and T. Kushida, Phys. Rev. B \textbf{5}, 3751 (1972).

\bibitem{Schl88}
M. Schlott, B. Elschner, M. Herrmann, and W. Assmus, Z. Phys. B \textbf{72}, 38 (1988).

\bibitem{Col87}
M. Coldea, H. Schaeffer, V. Weissenberger, and B. Elschner, Z. Phys. B \textbf{68}, 25 (1987).

\bibitem{Ott87} H.R. Ott and Z. Fisk, in
Handbook on the Physics and Chemistry of the Actinides,
edited by A. J. Freeman, G. H. Lander (Elsevier, Amsterdam, 1987).

\bibitem{Grewe91} N. Grewe and F. Steglich, in
Handbook on the Physics and Chemistry of the rare earths Vol. 14,
edited by K.A. Gschneider, L. Eyring (Elsevier, Amsterdam, 1991).

\bibitem{Abr70}
A. Abragam and B. Bleaney, Electron Paramagnetic Resonance of Transition Ions, Clarendon Press, Oxford (1970).

\bibitem{Moriya1979} T. Moriya, J. Magn. Magn. Mater. \textbf{14}, 1 (1979).

\bibitem{Ishigaki96}
A. Ishigaki and T. Moryia, J. Phys. Soc. Jpn. {\bf 65}, 3402 (1996).

\end{thebibliography}

\end{document}